\documentclass[prc,preprint,showpacs,preprintnumbers,amsmath,amssymb]{revtex4}

\usepackage[mathscr]{eucal}
\usepackage{graphicx}
\def\slr#1{\setbox0=\hbox{$#1$}           
   \dimen0=\wd0                                 
   \setbox1=\hbox{/} \dimen1=\wd1               
   \ifdim\dimen0>\dimen1                        
      \rlap{\hbox to \dimen0{\hfil/\hfil}}      
      #1                                        
   \else                                        
      \rlap{\hbox to \dimen1{\hfil$#1$\hfil}}   
      /                                         
   \fi}

\def\kp{k^{\,\prime}}
\def\kpsq{k^{\,\prime\,2}}
\def\ksq{k^2}

\def\myint#1{\!\int\!\!\frac{d^4\!{#1}}{(2\pi)^4}\,}

\def\gev#1{ GeV${}^{#1}$}
\def\be{\begin{eqnarray}}
\def\ee{\end{eqnarray}}

\def\cond#1{\langle\bar #1\rangle}

\renewcommand{\theequation}%
    {\arabic{section}.\arabic{equation}}
\makeatletter \@addtoreset{equation}{section} \makeatother

\begin{document}


\title{Description of Deconfinement at Finite Matter Density in a Generalized Nambu--Jona-Lasinio Model}

\author{Hu Li}
\author{C.M. Shakin}
 \email[email:]{casbc@cunyvm.cuny.edu}

\affiliation{%
Department of Physics and Center for Nuclear Theory\\
Brooklyn College of the City University of New York\\
Brooklyn, New York 11210
}%

\date{June, 2002}

\begin{abstract}
Recent years have seen extensive applications of the
Nambu--Jona-Lasinio (NJL) model in the study of matter at high
density. There is a good deal of interest in the predictions of
diquark condensation and color superconductivity, with suggested
applications to the study the properties of neutron stars. As the
researchers in this field note, the NJL model does not describe
confinement, so that one is limited to the study of the deconfined
phase, which may set in at several times nuclear matter density.
Recently, we have extended the NJL model to include a covariant
confinement model. Our model may be used to study the properties
of the full range of light mesons, including their radial
excitations, in the 1-3 GeV energy domain. Most recently we have
used our extended model to provide an excellent fit to the
properties of the $\eta(547)$ and $\eta^\prime(958)$ mesons and
their radial excitations. The mixing angles and decay constants
are given successfully in our model. In the present work our goal
is to include a phenomenological model of deconfinement at finite
matter density, using some analogy to what is known concerning
``string breaking" and deconfinement at finite temperature.
Various models may be used, but for this work we choose a specific
model for the density dependence of the parameters of our
confining interaction. We perform relativistic
random-phase-approximation (RPA) calculations of the properties of
the $\pi(138), K(495), f_0(980), a_0(980)$ and $K_0^*(1430)$
mesons and their radial excitations. In the model chosen for this
work, there are no mesonic states beyond about $2\rho_{NM}$, where
$\rho_{NM}$ is the density of nuclear matter. (The density for
deconfinement in our model may be moved to higher values by the
change of one of the parameters of the model.) This inability of
the model to support hadronic excitations at large values of the
density is taken as a signal of deconfinement. In addition to the
density dependence of the confining interaction, we use the
density-dependent quark mass values obtained in either the SU(2)
or SU(3)-flavor versions of the NJL model. We stress that other
assumptions for the density dependence of the confinement
potential, other than that used in this work, maybe considered in
future work, particularly if we are able to obtain further insight
in the dynamics of deconfinement at finite matter density.
\end{abstract}


\maketitle

\section{INTRODUCTION}

In recent years we have developed a generalized
Nambu--Jona-Lasinio (NJL) model that incorporates a covariant
model of confinement [1-5]. The Lagrangian of the model is \be
{\cal L}=&&\bar q(i\slr
\partial-m^0)q +\frac{G_S}{2}\sum_{i=0}^8[
(\bar q\lambda^iq)^2+(\bar qi\gamma_5 \lambda^iq)^2]\nonumber\\
&&-\frac{G_V}{2}\sum_{i=0}^8[
(\bar q\lambda^i\gamma_\mu q)^2+(\bar q\lambda^i\gamma_5 \gamma_\mu q)^2]\nonumber\\
&& +\frac{G_D}{2}\{\det[\bar q(1+\gamma_5)q]+\det[\bar
q(1-\gamma_5)q]\} \nonumber\\
&&+ {\cal L}_{conf}\,, \ee where the $\lambda^i(i=0,\cdots, 8)$
are the Gell-Mann matrices, with
$\lambda^0=\sqrt{2/3}\mathbf{\,1}$, $m^0=\mbox{diag}\,(m_u^0,
m_d^0, m_s^0)$ is a matrix of current quark masses and ${\cal
L}_{conf}$ denotes our model of confinement. Many applications
have been made in the study of light meson spectra, decay
constants, and mixing angles. In the present work we extend our
model to include a description of deconfinement at finite density.

There has been extensive application of the NJL model in the study
of matter at high density, with particular interest in diquark
condensation and color superconductivity [6-9]. These studies find
application in the study of neutron stars. The NJL model is the
model of choice, since little insight into the properties of
matter at finite density can be obtained in lattice simulations of
QCD. This problem is associated with the introduction of a
chemical potential, which makes the Euclidean-space fermion
determinant complex.

The use of the NJL model in the hadronic phase of matter is
limited, since the standard version of the model does not contain
a model of confinement [10-12]. It is clearly of value to extend
the NJL model so that one can study the full range of densities of
interest at this point in time. We are encouraged in this program
by recent results, obtained in lattice simulations of QCD with
dynamical quarks, that provide information on the temperature
dependence of the confining interaction [13]. It is generally
believed that the presence of matter will play a role similar to
that of finite temperature, with deconfinement taking place at
some finite density, which might be several times that of nuclear
matter. In the present work we make a specific assumption
concerning the density dependence of the confining field and then
calculate meson spectra in the presence of our density-dependent
confining interaction. We also take into account the density
dependence of the constituent quark masses, which is calculated in
the SU(2) or SU(3)-flavor version of the NJL model. As is well
known, the presence of matter leads to a reduction in the
magnitude of the quark vacuum condensates, which represents a
partial restoration of chiral symmetry in matter.

Our calculations of the properties of mesons in matter is made
using a covariant random-phase-approximation (RPA) formalism,
which we have developed for the study of mesons in vacuum [1-5].
The organization of our work is as follows. In Section II we
provide a short review of our treatment of Lorentz-vector
confinement in our generalized NJL model. In Section III we
describe the variation of the up, down and strange quark
constituent quark masses in matter. In Section IV we discuss some
recent work concerning the temperature dependence of the confining
interaction, as obtained in lattice simulations of QCD with
dynamical quarks. We also specify the density dependence of the
confining field that we use in this work in Section IV. In Section
V we comment upon the phenomenon of pion condensation. (In our
work we introduce a small density dependence of the coupling
constants of the NJL model to simulate effects that prevent the
formation of a pion condensate in nuclear matter.) In Section VI
we discuss our covariant RPA calculations of meson properties in
vacuum and indicate how these calculations are modified in matter.
Results of our RPA calculations of the properties of pseudoscalar
mesons in matter are presented in Section VII, while Section VIII
contains similar results for scalar mesons. In the case of scalar
mesons, we study the  $a_0(980), f_0(980)$, and $K_0^*(1430)$
mesons and their radial excitations. Finally, Section IX contains
some further discussion and conclusions.

\section{models of confinement}

There are several models of confinement in use. One approach is
particularly suited to Euclidean-space calculations of hadron
properties. In that case one constructs a model of the quark
propagator by solving the Schwinger-Dyson equation. By appropriate
choice of the interaction one can construct a propagator that has
no on-mass-shell poles when the propagator is continued into
Minkowski space. Such calculations have recently been reviewed by
Roberts and Schmidt [14]. In the past, we have performed
calculations of the quark and gluon propagators in Euclidean space
and in Minkowski space. These calculations give rise to
propagators which did not have on-mass-shell poles [15-18].
However, for our studies of meson spectra, which included
descriptions of radial excitations, we found it useful to work in
Minkowski space.

The construction of our covariant confinement model has been
described in a number of works [1-5]. We have made use of
Lorentz-vector confinement, so that the Lagrangian of our model
exhibits chiral symmetry. We begin with the form $V^C(r)=\kappa
r\mbox{exp}[-\mu r]$ and obtain the momentum-space potential via
Fourier transformation. Thus, \be V^C(\vec k-\vec
k\,^\prime)=-8\pi\kappa\left[\frac1{[(\vec k-\vec
k\,^\prime)^2+\mu^2]^2}-\frac{4\mu^2}{[(\vec k-\vec
k\,^\prime)^2+\mu^2]^3}\right]\,,\ee with the matrix form \be
\overline V{}\,^C(\vec k-\vec k\,^\prime)=\gamma^\mu(1)V^C(\vec
k-\vec k\,^\prime)\gamma_\mu(2)\,,\ee appropriate to
Lorentz-vector confinement. The potential of Eq. (2.1) is used in
the meson rest frame. We may write a covariant version of
$V^C(\vec k-\vec k^\prime)$ by introducing the four-vectors \be
\hat k^\mu=k^\mu-\frac{(k\cdot P)P^\mu}{P^2}\,, \ee  and \be \hat
k^{\prime\,\mu}=k^{\prime\,\mu}-\frac{(k^\prime\cdot
P)P^\mu}{P^2}\,. \ee Thus, we have \be V^C(\hat k-\hat
k\,^\prime)=-8\pi\kappa\left[\frac1{[-(\hat k-\hat
k\,^\prime)^2+\mu^2]^2}-\frac{4\mu^2}{[-(\hat k-\hat
k\,^\prime)^2+\mu^2]^3}\right]\,.\ee Originally, the parameter
$\mu=0.010$ GeV was introduced to simplify our momentum-space
calculations. However, in the light of the following discussion,
we can remark that $\mu$ may be interpreted as describing
screening effects as they affect the confining potential [13].

In our work, we found that the use of $\kappa=0.055$\gev2 gave
very good results for meson spectra. Here, $\kappa$ for the
Lorentz-vector potential is about one-fourth of the value of
$\kappa$ for Lorentz-scalar confinement. This difference arises
since the Dirac matrices $\gamma^\mu(1)\gamma_\mu(2)$ in Eq. (2.2)
give rise to a factor of 4 upon forming various Dirac trace
operations, so that the $\emph{effective}$ value of the string
tension is about the same in both Lorentz-scalar and
Lorentz-vector models of confinement.

The potential $V^C(r)=\kappa r\mbox{exp}[-\mu r]$ has a maximum at
$r=1/\mu$, at which point the value is $V_{max}=\kappa/\mu e=
2.023$ GeV. If we consider pseudoscalar mesons, which have $L=0$,
the continuum of the model starts at $E_{cont}=m_1+m_2+V_{max}$,
so that for $m_1=m_2=m_u=m_d=0.364$ GeV, $E_{cont}=2.751$ GeV. It
is also worth noting that the potential goes to zero for very
large r. Thus, there are scattering states whose lowest energy
would be $m_1+m_2$. However, barrier penetration plays no role in
our work. The bound states in the interior of the potential do not
communicate with these scattering states to any significant
degree. It is not difficult to construct a computer program that
picks out the bound states from all the states found upon
diagonalizing the random-phase-approximation Hamiltonian.

Bound states in the confining field may be found by solving the
equation for the mesonic vertex function shown in Fig. 1a.
Inclusion of the short-range NJL interaction leads to an equation
for the vertex shown in Fig. 1b. We will return to a consideration
of Fig. 1b when we discuss our covariant RPA formalism in Section
VI.

 \begin{figure}
 \includegraphics[bb=30 0 180 220, angle=90, scale=1.5]{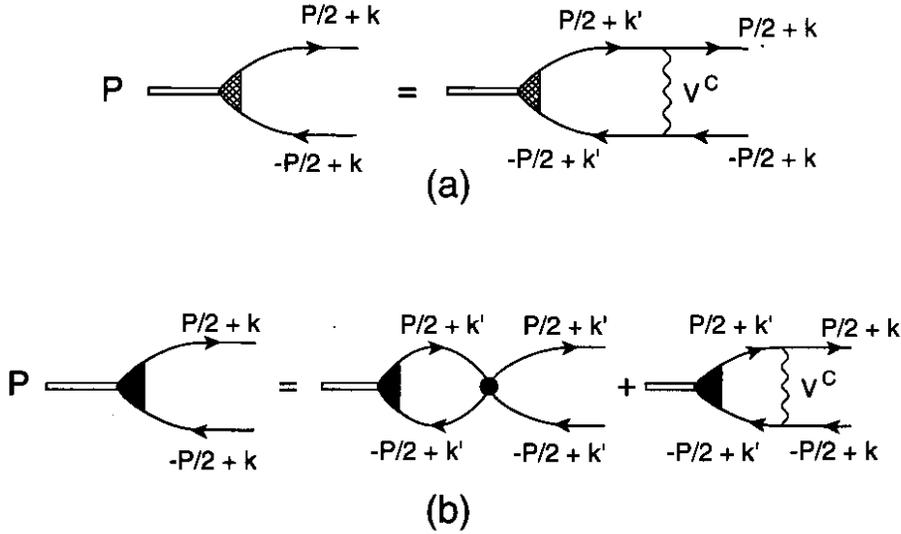}%
 \caption{a) Bound states in the confining field (wavy line) may be found by solving
 the equation for the vertex shown in this figure, b) Effects of both the confining
 field and the short-range NJL interaction (filled circle) are included when solving
 for the vertex shown in this figure.}
 \end{figure}

\section{calculation of constituent quark mass values}

In this Section we report upon our calculation of the density
dependence of the constituent quark masses of the up (or down) and
strange quarks. The role of confinement in the calculation of the
constituent mass was studied in an earlier work in which
calculations were made in Euclidean space [19]. The results were
similar to those obtained in Minkowski-space calculations in which
confinement was neglected and it is the latter calculations which
we discuss here.

The equations for the quark masses in the SU(3)-flavor NJL model
are [11] \be m_u=m_u^0-2G_S\langle\bar uu\rangle-G_D\langle\bar
dd\rangle\langle\bar ss\rangle\,,\\
m_d=m_d^0-2G_S\langle\bar dd\rangle-G_D\langle\bar
uu\rangle\langle\bar ss\rangle\,,\\
m_s=m_s^0-2G_S\langle\bar ss\rangle-G_D\langle\bar
uu\rangle\langle\bar dd\rangle\,,\ee where $\langle\bar uu\rangle,
\langle\bar dd\rangle$ and $\langle\bar ss\rangle$ are the quark
vacuum condensates. For example, with $N_c=3$, \be \langle\bar
uu\rangle=-4N_ci\myint k\frac{m_u}{\ksq-m_u^2+i\epsilon}\,.\ee If
this integral is evaluated in a Minkowski-space calculation, a
cutoff is used such that $|\vec k|\leq\Lambda_3$. Thus, \be
\langle\bar
uu\rangle=-4N_c\!\int^{\Lambda_3}\!\!\frac{d^3\!{k}}{(2\pi)^3}\,\frac{m_u}{2E_u(\vec
k)}\,,\ee etc. Here $E_u(\vec k)=\left[\vec
k{}^2+m_u^2\right]^{1/2}$.

For studies at finite density, we consider the presence of two
Fermi seas of up and down quarks with Fermi momentum $k_F$. We
also take $m_u^0=m_d^0$ and obtain the density-dependent
equations, with $\langle\bar uu\rangle_\rho=\langle\bar
dd\rangle_\rho$, \be m_u(\rho)=m_u^0-2G_S\langle\bar
uu\rangle_\rho-G_D\langle\bar
dd\rangle_\rho\langle\bar ss\rangle_\rho\,,\\
m_s(\rho)=m_s^0-2G_S\langle\bar ss\rangle_\rho-G_D\langle\bar
uu\rangle_\rho\langle\bar dd\rangle_\rho\,.\ee Equation (3.5) is
now replaced by \be \langle\bar
uu\rangle_\rho=-4N_c\!\left[\int_0^{\Lambda_3}\!\!\frac{d^3\!{k}}{(2\pi)^3}\,\frac{m_u(\rho)}
{2E_u(\vec
k)}-\int_0^{k_F}\!\!\frac{d^3\!{k}}{(2\pi)^3}\,\frac{m_u(\rho)}
{2E_u(\vec k)}\right]\,,\ee with $E_u(\vec k)=\left[\vec
k{}^2+m_u^2(\rho)\right]^{1/2}$. On the other hand, since we do
not consider a background of strange matter, we have \be
\langle\bar
ss\rangle_\rho=-4N_c\!\int_0^{\Lambda_3}\!\!\frac{d^3\!{k}}{(2\pi)^3}\,\frac{m_s(\rho)}
{2E_s(\vec k)}\,,\ee with $E_s(\vec k)=\left[\vec
k{}^2+m_s^2(\rho)\right]^{1/2}$.

We may argue that, with respect to our mean-field analysis, the
Fermi seas of up and down quarks yield contributions to the scalar
density that are similar to what would be obtained if the quarks
are organized into nucleons. One part of the argument is based
upon the well-known model-independent relation for the density
dependence of the condensate [20] \be \frac{\langle\bar
qq\rangle_\rho}{\langle\bar
qq\rangle_0}=\left(1-\frac{\sigma_N\rho}{f_\pi^2m_\pi^2}+\cdots\right)\,,\ee
where $\sigma_N$ is the pion-nucleon sigma term and $\rho$ is the
density of the matter. If we take $f_\pi=0.0942$ GeV,
$m_\pi=0.138$ GeV, $\rho_{NM}=(0.109\,\mbox{GeV})^3$ and
$\sigma_N=0.050$ GeV, we find a reduction of the condensate in
nuclear matter of 38\%, which is consistent with relativistic
models of nuclear matter [21, 22].

We now consider the corresponding relation for a quark gas of up
and down quarks, \be \frac{\langle\bar qq\rangle_\rho}{\langle\bar
qq\rangle_0}=\left(1-\frac{\sigma_q\rho_q}{f_\pi^2m_\pi^2}+\cdots\right)\,,\ee
where $\rho_q$ is the density of quarks ($\rho_q=3\rho$) and
$\sigma_q$ is a ``quark sigma term". We have shown in earlier work
[23] that $\sigma_q$ is in the range of 15-17 MeV, so that Eq.
(3.10) and (3.11) imply that quite similar mean fields are
generated by the quark gas and by nuclear matter.

In Table I and in Fig. 2, we show the results obtained when Eqs.
(3.8) and (3.9) are solved with $G_S=9.00$\gev{-2},
$G_D=-240$\gev{-5}, $\Lambda_3=0.631$ GeV, $m_u^0=0.0055$ GeV and
$m_s^0=0.130$ GeV. We note that the dependence of $m_u(\rho)$ on
density is approximately linear for $\rho/\rho_{NM}\leq2$, with a
32\% reduction in the value of $m_u(\rho)$ when
$\rho/\rho_{NM}=1$. Another point to note is that $m_s(\rho)$ is
density-dependent for finite values of $G_D$, since the
$\langle\bar ss\rangle$ condensate is modified by the coupling to
the up and down quark condensates via the 't Hooft interaction.
This coupling becomes less important as the up and down quark
condensates are reduced at increasing density. [See Fig. 2.]

 \begin{figure}
 \includegraphics[bb=0 0 400 220, angle=0, scale=1.5]{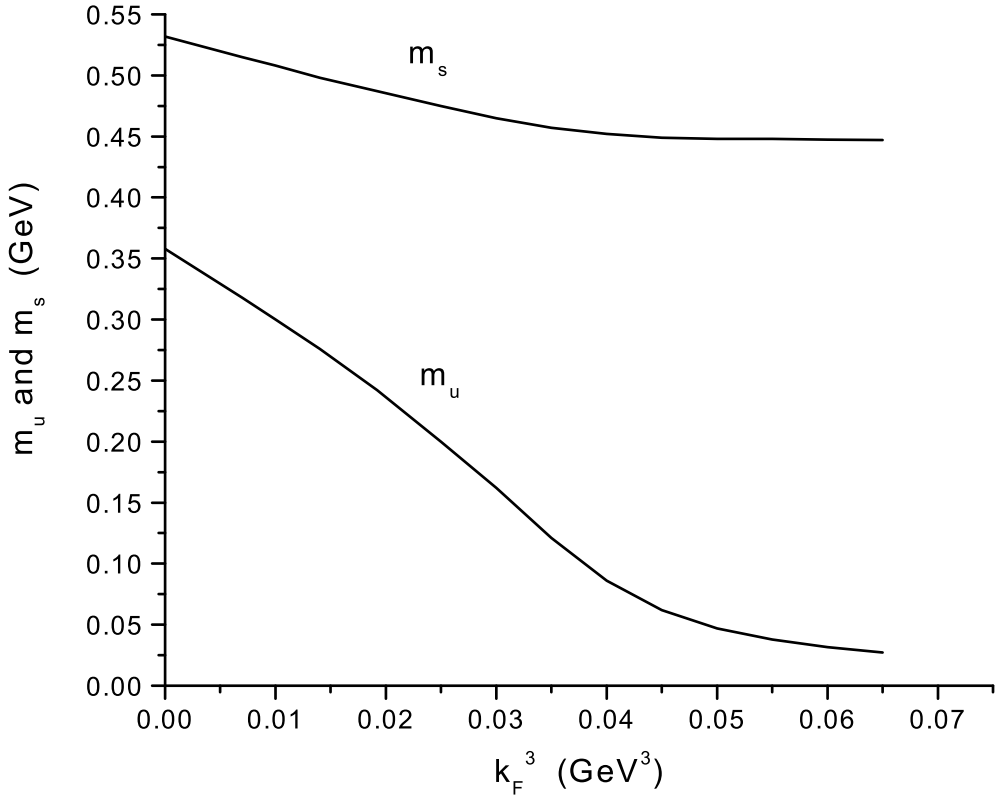}%
 \caption{The solution of Eqs. (3.6) and (3.7) for the density-dependent constituent
 quark masses, $m_u(\rho)=m_d(\rho)$ and $m_s(\rho)$ are shown.
 Here $G_S=9.00$\gev{-2}, $G_D=-240.0$\gev{-5}, $\Lambda_3=0.631$ GeV, $m_u^0=0.0055$ GeV
 and $m_s^0=0.130$ GeV.}
 \end{figure}

We have also considered the solution for the SU(2) version of the
above equations \be m_u(\rho)=m_u^0-2G_S\langle\bar
uu\rangle_\rho\,,\ee and have used the parameters specified in the
Klevansky review article [10], $G_S=10.15$\gev{-2}, $m_u^0=0.0055$
GeV and $\Lambda_3=0.631$ GeV. The results for $m_u(\rho)$ are
similar to that seen in Fig. 2, except that $m_u(0)=0.336$ GeV.
[See Fig. 3.] In this case, $m_u(\rho)$ is reduced by about 32\%
when $\rho=\rho_{NM}$.

\begin{table}
 \begin{tabular}{||@{\hspace{0.5cm}}
 c@{\hspace{0.5cm}}|@{\hspace{0.5cm}}c@{\hspace{0.5cm}}
 |@{\hspace{0.5cm}}c@{\hspace{0.5cm}}|@{\hspace{0.5cm}}c@{\hspace{0.5cm}}||}\hline\hline
 $k_F^3$     &$\rho/\rho_{NM}$               &$m_u(\rho)$                  &$m_s(\rho)$ \\
 (\gev{3})   &                               &[GeV]                        &[GeV]\\\hline\hline
 0.00        &0.00                           &0.358                        &0.532\\\hline
 0.007       &0.364                          &0.318                        &0.515\\\hline
 0.010       &0.521                          &0.300                        &0.508\\\hline
 0.0140      &0.729                          &0.276                        &0.498\\\hline
 0.0192      &1.00                           &0.242                        &0.487\\\hline
 0.025       &1.302                          &0.200                        &0.475\\\hline
 0.030       &1.562                          &0.162                        &0.465\\\hline
 0.035       &1.823                          &0.121                        &0.457\\\hline
 0.040       &2.083                          &0.0860                       &0.452\\\hline
 0.045       &2.343                          &0.0618                       &0.449\\\hline
 0.050       &2.604                          &0.0470                       &0.448\\\hline
 0.055       &2.864                          &0.0378                       &0.448\\\hline
 0.060       &3.125                          &0.0316                       &0.448\\\hline
 0.065       &3.385                          &0.0272                       &0.447\\\hline\hline

 \end{tabular}
 \vspace{1.2cm}
 \caption{Values of $m_u(\rho)$ and $m_s(\rho)$ obtained from the solution of Eqs. (3.6) and
 (3.7) are given for various values of the ratio $\rho/\rho_{NM}$. (Here, $k_F^3=0.0192$\gev3
 for nuclear matter, $m_u^0=0.0055$ GeV, $m_s^0=0.130$ GeV, $\Lambda_3=0.631$ GeV,
 $G_S=9.00$\gev{-2},  $G_D=-240.0$\gev{-5}.)}
 \end{table}

 \begin{figure}
 \includegraphics[bb=0 0 400 220, angle=-0.5, scale=1.5]{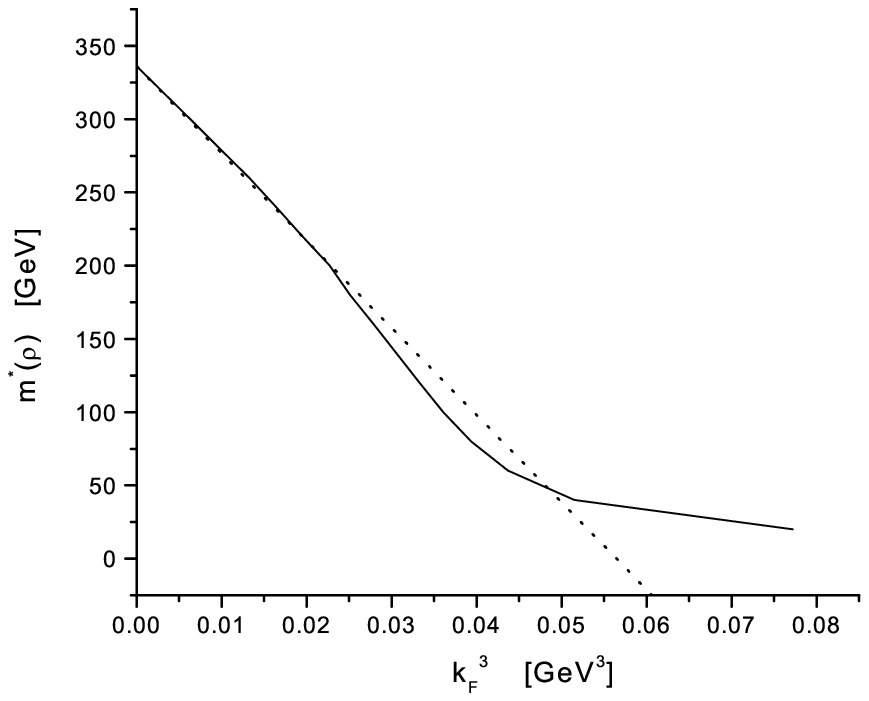}%
 \caption{The solution of Eq. (3.12) for $m_u(\rho)$ is shown. Here $G_S=10.15$\gev{-2},
 $m_u^0=0.0055$ GeV and $\Lambda_3=0.631$ GeV. (See Table V of Ref. [10].) The dashed line is a
 linear approximation to the result which we use for $\rho\leq2\rho_{NM}$.
 (Nuclear matter density
 corresponds to $k_F^3=0.0192$ \gev3.}
 \end{figure}

\section{density and temperature dependence of the confining field}

In part, our study has been stimulated by the results presented in
Ref. [13] for the temperature-dependent potential, $V(r)$, in the
case dynamical quarks are present. We reproduce some of the
results of that work in Fig. 4. There, the filled symbols
represent the results for $T/T_c=0.68, 0.80, 0.88$ and 0.94 when
dynamical quarks are present. This figure represents definite
evidence of ``string breaking", since the force between the quarks
appears to approach zero for $r > 1$ fm. This is not evidence for
deconfinement, which is found for $T=T_c$. Rather, it represents
the creation of a second $\bar qq$ pair, so that one has two
mesons after string breaking. Some clear evidence for string
breaking at zero temperature and finite density is reported in
Ref. [24].

 \begin{figure}
 \includegraphics[bb=0 0 250 350, angle=-90, scale=1.2]{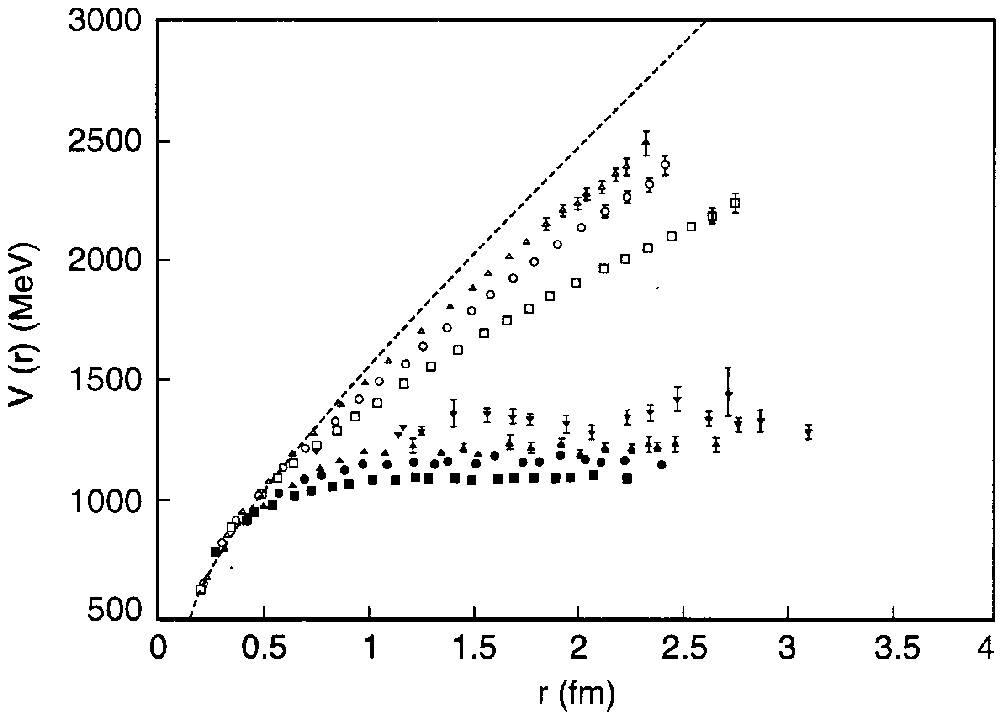}%
 \caption{A comparison of quenched (open symbols) and unquenched results (filled symbols) for
 the interquark potential at finite temperature [13]. The dotted line is the zero temperature
 quenched potential. Here, the symbols for $T=0.80T_c$ [open triangle], $T=0.88T_c$
 [open circle], $T=0.80T_c$ [open square], represent the quenched
 results. The results with dynamical fermions are given at $T=0.68T_c$ [solid downward-pointing
 triangle], $T=0.80T_c$ [solid upward-pointing triangle], $T=0.88T_c$ [solid circle],
 and $T=0.94T_c$ [solid square].}
 \end{figure}

In order to study deconfinement in our generalized NJL model, we
need to specify the interquark potential at finite density. We
start with our model that was described in Section II. In that
case we had $V^C(r)=\kappa r\mbox{exp}[-\mu r]$. For the model we
study in this work, we write \be V^C(r, \rho)=\kappa
r\mbox{exp}[-\mu(\rho) r]\ee and put \be
\mu(\rho)=\frac{\mu_0}{1-\left(\displaystyle\frac\rho{\rho_C}\right)^2}\,,\ee
with $\rho_C=2.25\rho_{NM}$ and $\mu_0=0.010$ GeV. With this
modification our results for meson spectra in the vacuum are
unchanged. Other forms than that given in Eqs. (4.1) and (4.2) may
be used. However, in this work we limit our analysis to the model
presented in these equations. The corresponding potentials for our
model of Lorentz-vector confinement are shown in Fig. 5 for
several values of $\rho/\rho_{NM}$.

 \begin{figure}
 \includegraphics[bb=0 20 400 220, angle=0, scale=1.5]{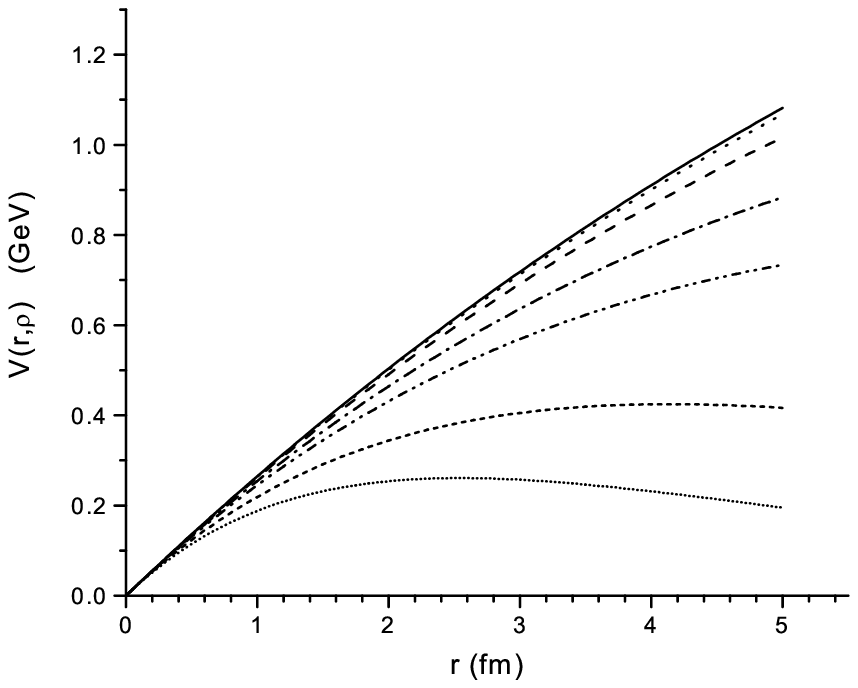}%
 \caption{Values of $V(r,\rho)$  are shown, where $V(r,\rho)=\kappa r\exp[-\mu(\rho)r]$
 and $\mu(\rho)=\mu_0/[1-(\rho/\rho_C)^2]$. Here $\rho_C=2.25\rho_{NM}$ and $\mu_0=0.010$ GeV.
 The values of $\rho/\rho_{NM}$ are 0.0 [solid line], 0.50 [dotted line], 1.0 [dash line],
 1.50 [dash-dot line]. 1.75 [dash-dot-dot line], 2.0 [short-dash line],
 and 2.1 [small dot line].}
 \end{figure}

We can see from Fig. 4 that, for $T=0.94T_c$, the use of dynamical
quarks leads to an approximately constant value of $V(r)=1000$ MeV
for larger $r$. If we perform a Fierz rearrangement of the
Lorentz-scalar potential to study pseudoscalar $q\bar q$ states,
one introduces a factor of 1/4, making the value at large $r$ to
be about 250 MeV. (See Eq. (B1) of Ref. [10].) However,
rearranging the Lorentz-vector potential to study pseudoscalar
$q\bar q$ states introduces a factor of 1. Now, let us consider
$\rho/\rho_{NM}=0.94(\rho_C/\rho_{NM})\simeq2.11$, and find the
maximum of our Lorentz-vector potential at that density from the
relation $V_{max}=\kappa/\mu(\rho)e$. Using our value for
$\mu(\rho)$ at $\rho/\rho_{NM}=2.11$, we obtain $V_{max}=0.227$
GeV. The value for the Lorentz-vector potential compares favorably
with the value of $V(r)$, for large $r$, quoted above. This result
suggests that, if the dynamics of chiral symmetry restoration and
deconfinement at finite temperature is somewhat analogous to the
deconfinement process at finite density, our use of
$\rho/\rho_{NM}=2.25$ may be a satisfactory choice.

\section{pion condensation and the choice of the parameters of the interaction}

It was suggested many years ago that the ground state of nuclear
matter might have an unusual structure due to presence of pionlike
excitations [25]. In finite nuclei such effects could imply
anomalous behavior in states with $J^\pi=0^-, 1^+, 2^-\ldots$,
etc. However, the nucleon-nucleon interaction is sufficiently
repulsive in the relevant channel so that pion condensation does
not take place at normal nuclear matter densities. That matter has
been discussed in Ref. [26]. A constant $g^\prime$ parametrizes
the strength of a nuclear force in the spin-isospin channel that
represents short-range correlation effects and exchange effects.
(See Eq. (5.11a) of Ref. [26].) The phenomenological value of
$g^\prime$, obtained from the study of nuclear excitations, is
sufficiently large so that pion condensation does not take place
until about three times nuclear matter density. (See Fig. 5.9 of
Ref. [26].)

In our work we will model the effects that prevent pion
condensation by introducing a density-dependent interaction for
the pionic states calculated in the NJL model. We write \be
G_\pi(\rho)=G_\pi(0)[1-0.087\rho/\rho_{NM}]\,,\ee where the second
term in Eq. (5.1) represents medium effects that reduce the pion
self-energy in matter. Here $G_\pi(0)$ is the linear combination
of $G_S$ and $G_D$ given on page 269 of Ref. [12], \be
G_\pi=G_S+\frac{G_D}2\langle\bar ss\rangle\,.\ee Equation (5.1)
represents our scheme for parametrizing the nuclear matter effects
that prevent pion condensation. In our calculations of pionlike
excitations we put $G_\pi(0)=13.49$\gev{-2}, and used a constant
values of $G_V=11.46$\gev{-2}. We may check that our choice of
$G_\pi(0)$ is reasonable by using Eq. (5.2) with
$G_S=11.84$\gev{-2} and $-180$\gev{-5}$\leq G_D\leq 240$\gev{-5}.
These values of $G_S$ and $G_D$ were obtained in our extensive
study of the eta mesons [1]. Thus, if we take $\langle\bar
ss\rangle=-(0.258\,\mbox{GeV})^3$ and $G_D=-190$\gev{-5}, we find
$G_\pi(0)=13.47$\gev{-2}. This analysis suggests that, once we fix
our parameters in the study of the eta mesons, we can then infer
the parameters needed for our study of the pion in vacuum.

For this work, in our study of the kaon, we use
$G_K(0)=13.07$\gev{-2} and $G_V=11.46$\gev{-2}. Note that [12] \be
G_K(0)=G_S+\frac{G_D}2\cond{dd}_0\,.\ee If we take
$G_S=11.84$\gev{-2}, $G_D=-190$\gev{-5} and
$\cond{uu}=-(0.240\,\mbox{GeV})^3$, we find
$G_K(0)=13.15$\gev{-2}, which is close to the value of
$G_K(0)=13.07$\gev{-2} used in our calculations. In our work we
have used \be G_K(\rho)=G_K(0)[1-0.087\rho/\rho_{NM}]\,.\ee In the
case of the kaon, about 40\% of the assumed density dependence of
$G_K(\rho)$ may be attributed to the density dependence of
$\cond{uu}_\rho$ or $\cond{dd}_\rho$. We may consider the relation
\be G_K(\rho)=G_S(\rho)+\frac{G_D}2\cond{dd}_\rho\,,\ee and use a
somewhat smaller reduction of $G_S(\rho)$ for the kaon than that
used for the pion in Eq. (5.5), since the reduction of
$\cond{uu}_\rho$ or $\cond{dd}_\rho$ in matter effectively reduces
the interaction strength.

In the absence of $a_0-f_0$ coupling we have
$G_{33}^S=G_{a_0}=G_S-(G_D/2)\cond{ss}$ [12]. If we again put
$G_S=11.84$\gev{-2}, $G_D=-190$\gev{-5}, and
$\cond{ss}=-(0.258\,\mbox{GeV})^3$, we have
$G_{a_0}=10.21$\gev{-2}, which places the $a_0(980)$ at 1.13 GeV.
However, in the case of the scalar mesons there exist significant
contributions to the interaction from processes that describe the
scalar meson decay to two-meson channels. An extended discussion
of these effects was given in an early work on scalar mesons [27].
In the case of the $f_0(980)$ we presented a discussion of such
terms as they affect the energy predicted for the $f_0(980)$ in
Ref. [28].

In order to take into account these effects, which are not
included in our RPA calculations, we increase the value of the
$a_0$ coupling constant to $G_{a_0}=13.10$\gev{-2}. That has the
effect of moving the $a_0(980)$ mass down to 980 MeV.

We also introduce some density dependence of the interaction to
avoid an ``$a_0$ condensate", which would otherwise take place at
$\rho=1.75\rho_{NM}$, if we use
$m_u(\rho)=m_d(\rho)=0.0055+0.3585(1-0.4\rho/\rho_{NM})$. Thus, we
use $G_{a_0}(\rho)=G_{a_0}(0)[1-0.045\rho/\rho_{NM}]$ when we
allow for the rapid decrease in the value of $m_u(\rho)=m_d(\rho)$
given by the above expression. It is possible that the small
reduction of $G_{a_0}(\rho)$ in matter given above has it origin
in a somewhat smaller attraction generated at the larger densities
by the real part of the polarization operator that describes decay
to the two-meson channels [27, 28]. We will provide further
details of our treatment of the scalar mesons in Section VIII.

\section{random phase approximation for mesonic excitations}

In this work we report upon covariant random-phase-approximation
(RPA) calculations of meson spectra in vacuum and in dense matter.
Before writing the equations of our model, it is worth discussing
some properties of RPA calculations made for many-body systems
[29, 30]. For example, such calculations have been performed to
study excited states of nuclei. In the RPA one usually does not
attempt to construct the wave function of the ground state.
Rather, one considers amplitudes of particle-hole operators taken
between the excited state and the ground state. The dominant
amplitude usually involves the creation of a hole in the ground
state and the creation of a particle in what are predominantly
unoccupied states. Smaller amplitudes are found if one destroys a
hole in the ground state and destroys a particle in the
predominantly unoccupied states. These smaller amplitudes are only
nonzero, if one allows for correlations in the ground state.

Such RPA calculations are particularly important for states that
are collective with respect to matrix elements of electromagnetic
transition operators, for example. In hadron physics the most
``collective state" is the $\pi(138)$. In this case the ``large"
and ``small" components of the wave function, in the sense of the
RPA, are comparable in magnitude and approach equality in
magnitude as one approaches the chiral limit, when
$m_\pi\rightarrow0$.

Another important feature of RPA calculations is that they may be
considered as an investigation of the properties of small
oscillations about the ground state. Thus, if one obtains an
imaginary energy value for the ground state, one infers that the
ground state is unstable. A new ground state must be constructed
that will yield real eigenvalues. (Note that imaginary eigenvalues
may be obtained, since the RPA Hamiltonian is not Hermitian.)

There is a strong analogy that can be made between the
particle-hole RPA calculations described above and the calculation
of mesonic excitations. For example, a ``hole" in the ground state
(the vacuum) is an antiquark, while the particle state is the
quark. If we perform relativistic RPA calculations for the pion
and its radial excitations, an imaginary energy calculated for the
pion is a signal of pion condensation.

Random-phase-approximation equations may be derived using the
vertex equation of Fig. 1b. The RPA equations for the study of the
pion, kaon, and eta mesons were derived in Ref. [1]. In the case
of the pion and kaon we include pseudoscalar---axial-vector
coupling. The most complex case is that of the eta mesons which,
in addition to pseudoscalar---axial-vector coupling, involves
singlet-octet coupling in the flavor sector.

In this work we only record the equations in the simplest example,
that of RPA calculations for the $a_0$ mesons [31]. In this case
the large component is denoted as $\phi^+(k)$, while the small
component is $\phi^-(k)$. These functions are found to satisfy
coupled equations for mesons in vacuum: \be
2E_u(k)\phi^+(k)+\int\!
d\kp\,[H_C(k,\kp)+H_{NJL}(k,\kp)]\phi^+(\kp)\\\nonumber+\int\!
d\kp \,H_{NJL}(k,\kp)\phi^-(\kp)=P^0\phi^+(k)\,,\ee \be
-2E_u(k)\phi^-(k)-\int\!
d\kp\,[H_C(k,\kp)+H_{NJL}(k,\kp)]\phi^-(\kp)\\\nonumber-\int\!
d\kp \,H_{NJL}(k,\kp)\phi^+(\kp)=P^0\phi^-(k)\,,\ee where
$E_u(k)=[\vec k\,{}^2+m_u^2]^{1/2}$, \be
H_C(k,\kp)=-\frac1{(2\pi)^2}\frac{[2V_0^C(k,\kp)k^2\kpsq+k\kp
V_1^C(k,\kp)]}{E_u(k)E_u(\kp)}\,,\ee and \be
H_{NJL}=\frac{8N_c}{(2\pi)^2}\frac{\ksq\kpsq
G_{a_0}e^{-\ksq/2\alpha^2}e^{-\kpsq/2\alpha^2}}{E_u(k)E_u(\kp)}\,.\ee
In Eq. (6.3) we have introduced \be
V_l^C(k,\kp)=\frac12\int_{-1}^1\!dx\, P_l(x)V^C(\vec k-\vec
k{}^\prime)\,.\ee Here, $x=\mbox{cos}\theta$ and $P_l(x)$ is a
Legendre function. The terms exp[$-\ksq/2\alpha^2$] and
exp[$-\kpsq/2\alpha^2$] are regulators with $\alpha=0.605$ GeV.

In order to solve these equations in the presence of matter, we
replace $m_u$, $G_{a_0}$ and $\mu_0$ by $m_u(\rho)$,
$G_{a_0}(\rho)$ and $\mu(\rho)$. (Recall that
$\mu(\rho)=\mu_0/[1-(\rho/\rho_C)^2]$.) In our calculation for the
$a_0$ states we have taken
$m_u(\rho)=m_u^0+0.3585\,\mbox{GeV}\,[1-0.4\rho/\rho_{NM}]$ and
$G_{a_0}(\rho)=G_{a_0}(0)[1-0.045\rho/\rho_{NM}]$, with
$m_u^0=0.0055$ GeV. As an alternative, the mass values for
$m_u(\rho)=m_d(\rho)$ may be taken from Table I.

\section{results of numerical calculations: pseudoscalar mesons}

The choice of the parameters in the case of the pion and its
radial excitations was discussed in Section V. We use
$G_\pi(\rho)=G_\pi(0)[1-0.087\rho/\rho_{NM}]$ and
$m_u(\rho)=m_d(\rho)=0.0055+0.3585[1-0.4\rho/\rho_{NM}]$ with
$G_\pi(0)=13.49$\gev{-2} and $G_V=11.46$\gev{-2}. Also,
$\mu(\rho)=\mu_0/[1-(\rho/\rho_C)^2]$ with $\mu_0=0.010$ GeV and
$\rho_C=2.25\rho_{NM}$.

The results of our calculations are shown in Fig. 6. At $\rho=0$,
the first radial excitation of the pion is found at 1.319 GeV. The
large number of states above 1.3 GeV have wave functions that are
dominated by either the $\gamma_5$ or $\gamma_0\gamma_5$ vertex.
The pion wave function has mainly a $\gamma_5$ vertex structure,
with a small admixture of the $\gamma_0\gamma_5$ vertex. (The
axial-vector part of the wave function makes a significant
contribution in the calculation of the pion decay constant,
$f_\pi$.)

 \begin{figure}
 \includegraphics[bb=0 50 280 280, angle=0, scale=1.5]{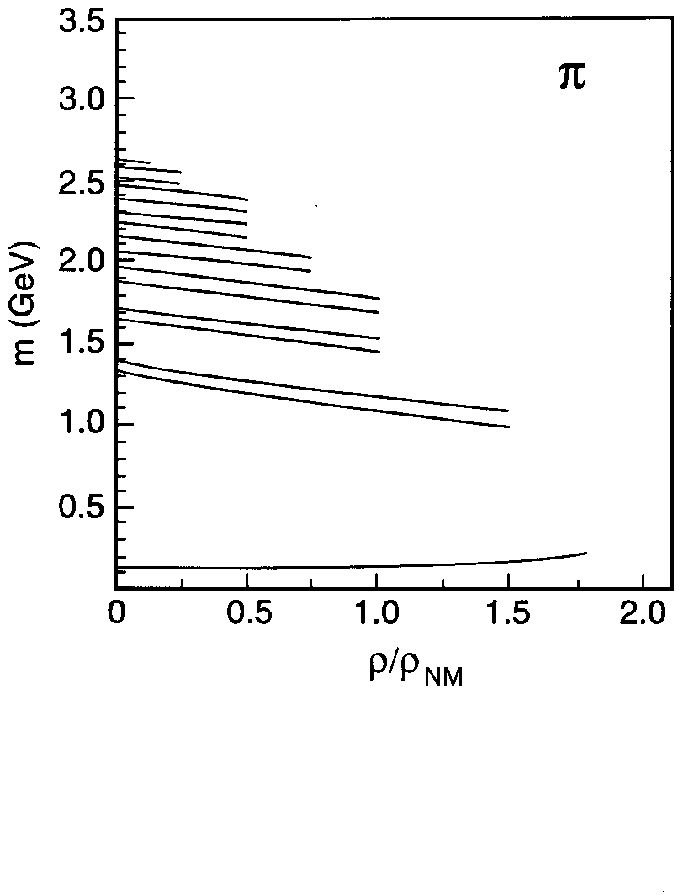}%
 \caption{The mass values for the pion and its radial excitations are presented as a function of
 the density of matter. Here, $G_\pi(\rho)=G_\pi(0)[1-0.087\rho/\rho_{NM}]$
 and $m_u(\rho)=m_d(\rho)=m_u^0+0.3585\,\mbox{GeV}[1-0.4\rho/\rho_{NM}]$,
 with $m_u^0=0.0055$ GeV. We use $G_\pi(0)=13.49$\gev{-2}, $G_V=11.46$\gev{-2}
 and $\mu=\mu_0/[1-(\rho/\rho_C)^2]$, with $\mu_0=0.010$ GeV and $\rho_C=2.25\rho_{NM}$.}
 \end{figure}

It may be seen from the figure, that with the reduction of the
value of the constituent mass and of the confining field with
increasing values of $\rho/\rho_{NM}$, the radial excitations that
appear as bound states become fewer in number. Beyond
$\rho/\rho_{NM}=1.50$ only the nodeless pion wave function is
bound and that state is no longer supported beyond
$\rho/\rho_{NM}\simeq1.80$. That represents the beginning of the
deconfined phase in the case of the pion for the model introduced
in this work.

Somewhat similar behavior is found for the kaon and its radial
excitations, as may be seen in Fig. 7. Here we have used the mass
values given in Table I and
$G_K(\rho)=G_K(0)[1-0.087\rho/\rho_{NM}]$ with
$G_K(0)=13.07$\gev{-2} and $G_V=11.46$\gev{-2}. Again we see only
a small increase of the mass of the nodeless state, the pseudo
Goldstone boson, as $\rho/\rho_{NM}$ is increased. We again find
deconfinement for $\rho/\rho_{NM}>1.8$. The density dependence of
$G_K(\rho)$ is taken to be the same as in the case of the pion.
However, in this case, we have noted previously that about 40\% of
the reduction of $G_K(\rho)$ with increasing density may be
ascribed to the density dependence of the up and down quark
condensates, $\cond {uu}_\rho$ and $\cond {dd}_\rho$. The
calculation of the density dependence of the coupling constants in
our model is a major undertaking and is beyond the scope of this
work.

 \begin{figure}
 \includegraphics[bb=0 0 280 220, angle=0, scale=1.5]{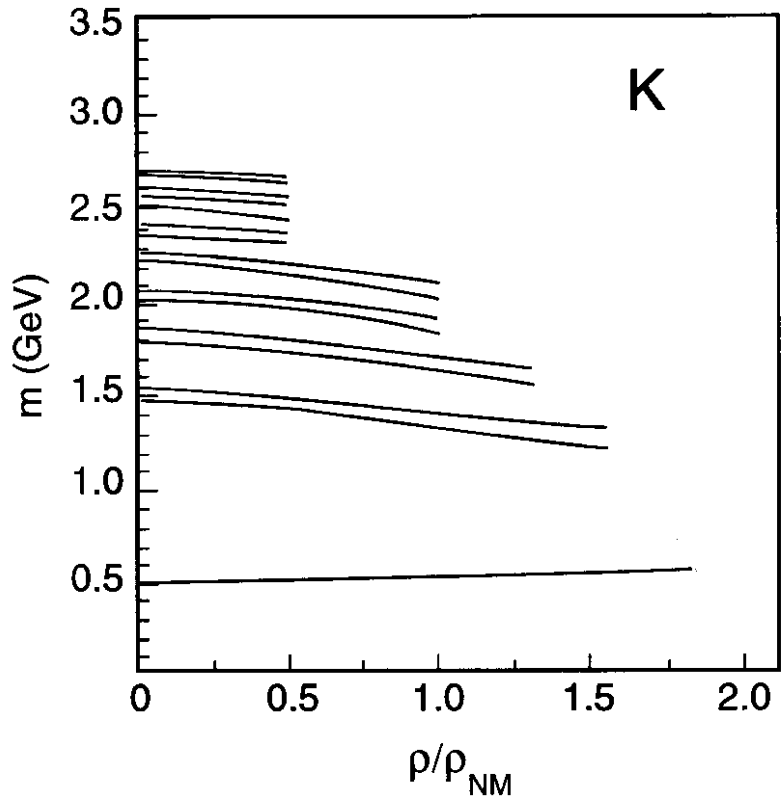}%
 \caption{Mass values of the $K$ mesons are shown as a function of the density of matter.
 Here we use $G_K(0)=13.07$\gev{-2}, $G_K(\rho)=G_K(0)[1-0.087\rho/\rho_{NM}]$,
 $G_V=11.46$\gev{-2} and $\mu=\mu_0/[1-(\rho/\rho_C)^2]$, with $\mu_0=0.010$ GeV
 and $\rho_C=2.25\rho_{NM}$. The mass values given in Table I are used.}
 \end{figure}

\section{results of numerical calculations: scalar mesons}

We have recently discussed the properties of the $f_0(980)$,
giving particular attention to the role of the polarization
diagrams that describe the decay of the $f_0$ mesons to the
$\pi\pi$ or $K\bar K$ channels [28]. (See Fig. 2 of Ref. [28].)
However, when we diagonalize the RPA Hamiltonian we do not take
those terms into account. Calculations of such effects are more
easily made if we construct a quark-antiquark $T$ matrix. For a
single channel example we may write \be t(p^2)=-\frac
G{1-GJ(p^2)}\,,\ee where $G$ is the appropriate coupling constant
for that channel and $J(p^2)$ is the corresponding vacuum
polarization operator. In our model $J(p^2)$ is calculated with
the confining vertex function that appears in Fig. 1a as a
crosshatched region. (See Fig. 1 of Ref. [28].) The resulting
$J(p^2)$ is a real function, which is singular at the values of
$p^2$ for which there is a bound state in the confining field. If
we include polarization diagrams that describe coupling to
two-meson decay channels, Eq. (8.1) is modified to read \be
t(p^2)=-\frac
G{1-G[J(p^2)+\mbox{Re}K(p^2)+i\mbox{Im}K(p^2)]}\,.\ee The
calculation of $J(p^2)$ and $K(p^2)$ has been extensively
discussed in our earlier work. In the case of the scalar mesons,
inclusion of $\mbox{Re}K(p^2)$ can move the mass of the
lowest-energy state down by about 70-100 MeV [27, 28].

In the case of the $a_0(980)$, the use of $G_S$ and $G_D$
determined in our study of the eta mesons places the $a_0(980)$ at
1.13 GeV. In the present work we have increased the coupling
constant from $G_{a_0}=10.21$\gev{-2} to $G_{a_0}=13.10$\gev{-2}
to move the lowest $a_0$ state down to 980 MeV. That creates a
problem of ``$a_0$ condensation" which we avoid by taking
$G_{a_0}(\rho)=G_{a_0}(0)[1-0.045\rho/\rho_{NM}]$. One may
speculate that the effects that increase the effective coupling
strength from $G_{a_0}=10.21$\gev{-2} to $G_{a_0}=13.10$\gev{-2}
have some density dependence that reduces the induced attraction
at the higher densities.

In Fig. 8 we show our results for the $a_0$ mesons. There we see
deconfinement at about $\rho=2.0\rho_{NM}$ which is a slightly
larger value of the density than that found for the other mesons
studied in this work. However, the behavior of the lowest $a_0$
state with increasing density is made somewhat uncertain because
of our lack of knowledge of the appropriate form for
$G_{a_0}(\rho)$.

 \begin{figure}
 \includegraphics[bb=0 60 400 400, angle=0, scale=1]{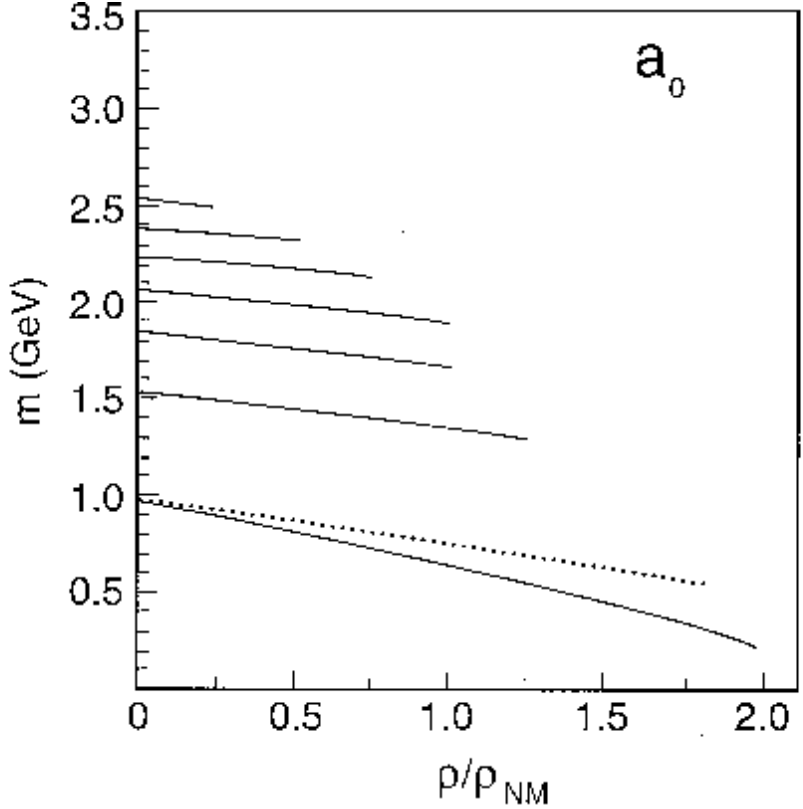}%
 \caption{Mass values for the $a_0$ mesons are given as a function of the matter density.
 Here, we have used $G_{a_0}(0)=13.10$\gev{-2} and
 $G_{a_0}(\rho)=G_{a_0}(0)[1-0.045\rho/\rho_{NM}]$. We have
 used $m_u=m_u^0+0.3585\,\mbox{GeV}[1-0.4\rho/\rho_{NM}]$
 with $m_u^0=0.0055$ GeV. The dotted line results, if we put
 $G_{a_0}(\rho)=G_{a_0}(0)[1-0.087\rho/\rho_{NM}]$ and use the mass values of
 Table I. The dotted curve is similar to the curve for the
 $a_0$ mass given in Ref. [31]. The curves representing the masses of
 the radial excitations are changed very little when we use the second
 form for $G_{a_0}(\rho)$ given above.}
 \end{figure}

For our study of the $f_0$ mesons we work in a singlet-octet
representation and use the coupling constants
$G_{00}^S=14.25$\gev{-2}, $G_{88}^S=10.65$\gev{-2} and
$G_{08}^S=G_{80}^S=0.4953$\gev{-2}. This choice yields 980 MeV for
the mass of the $f_0(980)$. The fact that $G_{00}^S>G_{88}^S$ is a
feature of the 't Hooft interaction and leads to the $f_0(980)$
being mainly a singlet state [28]. (For the $\eta(547)$ the
behavior of the 't Hooft interaction is such that
$G_{88}^S>G_{00}^S$ [12, 28] and, therefore, the $\eta(547)$ is
predominantly a flavor octet meson [1].)

In our study of the $f_0$ mesons at finite density we use the mass
values of Table I and do not introduce any density dependence for
$G_{00}^S$, $G_{88}^S$ and $G_{08}^S$. The results of our
calculation are shown in Fig. 9. Since the $f_0(980)$ has a
significant $\bar ss$ component, the mass value only decreases
slowly, with a value of 700 MeV for the lowest $f_0$ state at
$\rho/\rho_{NM}=1.82$, where deconfinement sets in.

 \begin{figure}
 \includegraphics[bb=0 50 280 270, angle=0, scale=1.5]{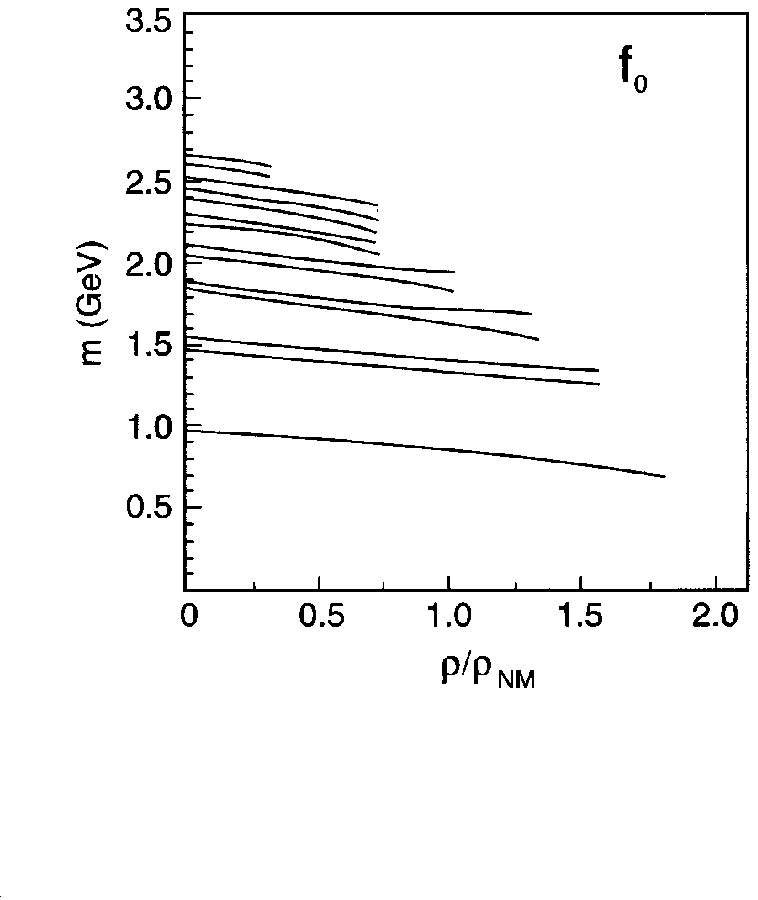}%
 \caption{The figure shows the mass values of the $f_0$ mesons as a
 function of density. The mass values for the quarks are taken from
 Table I. In a singlet-octet representation, we have used the
 constants $G_{00}^S=14.25$\gev{-2}, $G_{08}^S=10.65$\gev{-2} and $G_{88}^S=0.4953$\gev{-2}.
 Deconfinement takes place somewhat above $\rho=1.8\rho_{NM}$.
 Here $\mu=\mu_0/[1-(\rho/\rho_C)^2]$
 with $\mu_0=0.010$ GeV and $\rho_C=2.25\rho_{NM}$.}
 \end{figure}

In Ref. [28] we provide a discussion of the $T$ matrix for the
singlet-octet channels. There the role of $K_{00}^S(p^2)$,
$K_{08}^S(p^2)$ and $K_{88}^S(p^2)$ in lowering the energy
predicted for the $f_0(980)$ is discussed in some detail.

Our results for the energy levels of the $K_0^*$ mesons are given
in Fig. 10. In this case we use a constant value for
$G_{K_0^*}=10.25$\gev{-2}. The results are hardly modified if we
allow for a small density dependence of $G_{K_0^*}$. Since the
$K_0^*$ mesons contain a strange quark, the density dependence of
their energies is not as marked as that of the $a_0$ mesons which
only contain up and down quarks in our model. In that regard, the
behavior of the $K_0^*$ mesons is more like that of the $f_0$
mesons, which have some strange quark content. Again we see
deconfinement for $\rho>1.8\rho_{NM}$.

 \begin{figure}
 \includegraphics[bb=0 50 280 270, angle=0, scale=1.5]{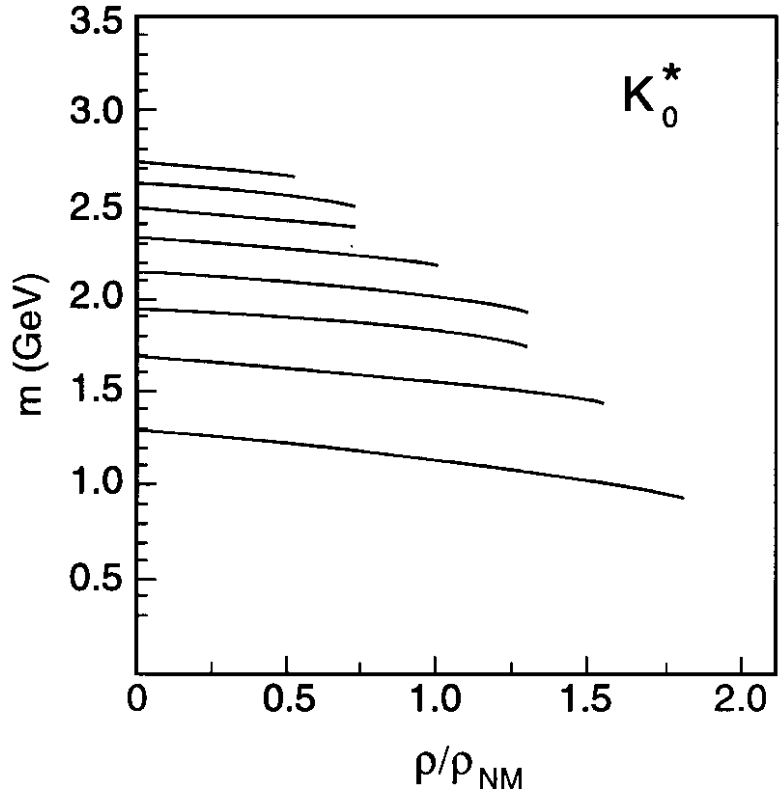}%
 \caption{The figure shows the mass values obtained for the $K_0^*$ mesons as a
 function of density. Here we use a constant $G_{K_0^*}=10.25$\gev{-2}.
 Deconfinement takes place somewhat above $\rho=1.8\rho_{NM}$. The quark mass values
 were taken from Table I.}
 \end{figure}

\section{discussion}

We originally chose $\rho_C=2.25\rho_{NM}$, since the curve in
Fig. 2 that shows the values of $m_u(\rho)$ seemed to change its
behavior at about $k_F^3=0.045$\gev3, which corresponds to
$\rho\simeq2.3\rho_{NM}$. We can attempt to see if that is a
reasonable choice by noting that ``string breaking" should occur
when the energy of the extended string is equal to the energy of
the lowest two-meson state that can be formed when the string
breaks. Therefore, we may write $V_{max}=m_1+m_2$, where $m_1+m_2$
are the masses of the mesons in the final state. We then use
$V_{max}=\kappa/\mu(\rho)e$ to find a value $\mu(\rho)$ and obtain
$\rho/\rho_C$ from the expression
$\mu(\rho)=\mu_0/[1-(\rho/\rho_C)^2]$. We then put
$\rho_C=2.25\rho_{NM}$ and calculate the value of $\rho/\rho_{NM}$
where we might expect string breaking. We consider the final
states $\pi\pi$, $\pi K$, $\pi\eta$ and $K\bar K$. The
corresponding values of $\rho/\rho_{NM}$ are 2.09, 1.86, 1.83, and
1.61 for $\rho_C=2.25\rho_{NM}$. Note that the $K(495)$ and
$K_0^*(1430)$ mesons can break up into the $\pi K$ system, while
the $a_0(980)$ is strongly coupled to the $\pi\eta$ channel. The
$f_0(980)$ is coupled both to the $\pi\pi$ and $K\bar K$ channels.
On the whole, the values of $\rho/\rho_{NM}$ calculated above are
generally consistent with the value of that quantity that leads to
deconfinement in our model. That result tends to suggest that, for
light mesons, the density that leads to string breaking may be
similar to the density for deconfinement. (In general, however,
these processes are distinct and further studies would be needed
to see if string breaking and deconfinement are related at finite
density.) We may suggest that, if the initial meson is of the same
type as the mesons that appear upon string breaking, it becomes
reasonable to suggest that the instability of the initial mesons
is also felt by the final state mesons, giving rise to the
relation of string breaking and deconfinement suggested above for
light mesons.

A comprehensive discussion of meson properties at finite
temperature and density has been presented by Lutz, Klimt and
Weise [32]. Since those authors did not include a model of
confinement, they were able to calculate values of the meson
masses for large values of the density. Their Fig. 8 shows the
calculated masses of the nodeless pion, $f_0$ and $a_0$ mesons for
$0\leq\rho/\rho_{NM}\leq 3.5$. They also give the result for an
$f_0^\prime$ excitation. (The $f_0$ and $f_0^\prime$ exhibit
singlet-octet mixing.) Compared to our results, their value of the
$f_0$ mass falls more rapidly than ours, becoming degenerate with
the pion mass at about $\rho/\rho_{NM}=3$. On the other hand, the
mass of the $a_0$ in their work is about 600 MeV at
$\rho/\rho_{NM}=2$. They are able to derive systematic low-density
expansions for various quantities which provide important insight
into the results obtained in numerical studies. They also show
that effects due to finite quasiparticle size are important in
stabilizing the density and temperature dependence of the pion
mass. The main deficiency of their work is the absence of a model
of confinement. Therefore, we believe our work provides a natural
extension of the work reported in Ref. [32].

It is worth noting that deconfinement takes place in our model at
about $\rho=1.8\rho_{NM}$, while the confining potential goes to
zero at $\rho=\rho_C=2.25\rho_{NM}$. That suggests that the
specific form we have chosen for the density dependence,
$\mu(\rho)=\mu_0/[1-(\rho/\rho_C)^2]$, is not particularly
important. What is more important is the behavior of our confining
potential, $V^C(r, \rho)$, shown in Fig. 5. There, we see that the
potential still has a substantial magnitude at
$\rho=1.75\rho_{NM}$ and $\rho=2.10\rho_{NM}$.

Since the analysis of Ref. [32] is made in the absence of a model
of confinement, many analytic results can be obtained for the
behavior of various quantities when small changes in density and
temperature are considered. Indeed, the work of that reference
provides some support for our treatment of the pion and kaon. It
is shown that the Goldstone boson remains at zero mass in the
chiral limit as long as the system remains in the Goldstone-Nambu
mode of symmetry breaking. For finite current quark masses, we
quote the result given in Eq. (5.6) of Ref. [32] for $T=0$, \be
\frac{dm_\pi^2}{m_\pi^2}=\left(1-2m_u^2\langle
r_S^2\rangle\right)\frac{d\cond{uu}}{\cond{uu}}\,.\ee Here, $r_S$
is the quasiparticle radius. That quantity is defined in terms of
the form factor $F_S(\vec p-\vec p\,{}^\prime)$ that appears in
the matrix element of the u-quark scalar density \be \langle
u(\vec p\,{}^\prime)|\,\bar uu(0)|\,u(\vec p)\rangle=F_S(\vec
p-\vec p\,{}^\prime)\bar u(p\,{}^\prime)u(p)\,.\ee In Eq. (9.2)
$u(\vec p)$ denotes the Dirac spinor of a constituent u quark with
four-momentum $p$. The scalar mean-squared radius is then \be
\left.\langle r_S^2\rangle=6\frac d{dq^2}\ln
F_S(q^2)\right|_{\,q^2=0}\,.\ee (See Eq. (A.7) of Ref. [32] for an
explicit expression for $\langle r_S^2\rangle$ in terms of the
parameters of the NJL model.) With the well-known relation [20]
\be
\frac{d\cond{uu}}{\cond{uu}}=-\frac{\sigma_N\rho}{m_\pi^2f_\pi^2}\,,\ee
Eq. (9.1) becomes \be dm_\pi^2=-\left(1-2m_u^2\langle
r_S^2\rangle\right)\frac{\sigma_N\rho}{f_\pi^2}\,.\ee If one
ignores the quasiparticle size, one has
$dm_\pi^2=-(\sigma_N\rho/f_\pi^2)$ [33, 34], which implies pion
condensation at a critical density
$\rho_{crit}=f_\pi^2m_\pi^2/\sigma_N=(0.148\,\mbox{GeV})^3$, which
is about 2.5 $\rho_{NM}$.

The second term in Eq. (9.5) works against condensation. With
$m_u=0.364$ GeV and $r_S=0.40$ fm [32] one finds that $\delta
m_\pi^2$ increases slowly with increasing density, as born out by
the calculations reported in Ref. [32]. Our choice of
$G_\pi(\rho)=G_\pi(0)[1-0.087\rho/\rho_{NM}]$ reproduces the
almost constant value of $m_\pi$. We see that the
density-dependent term in $G_\pi(\rho)$ plays a similar role in
our model as that played by the second term in Eq. (9.5).

We have some confidence in our treatment of the pion and kaon at
finite density. We recall that we were able to find satisfactory
values of $G_\pi(0)$ and $G_K(0)$ using the values of $G_S$ and
$G_D$ obtained in our study of the eta mesons [1]. Therefore, our
work provides a unified approach for the nonet of pseudoscalar
mesons in the presence of a model of confinement.

Since confinement is important for the $a_0(980)$ and $f_0(980)$
mesons, it is uncertain whether the results of Ref. [32] for the
properties of these mesons can be trusted. These mesons are in the
continuum of the NJL model without confinement and various
assumptions need to be made as to how the formalism is to be
applied. For a small increase in density, the mass of the $a_0$ in
our model and in Ref. [32] are similar. For the larger values of
density, the use of
$G_{a_0}(\rho)=G_{a_0}(0)[1-0.045\rho/\rho_{NM}]$ in model leads
to a rather small mass for the $a_0$ for $\rho\sim2\rho_{NM}$.
[See Fig. 8.]

Our treatment of the $a_0$ mesons is less satisfactory than that
of $\pi$ and $K$ mesons, since coupled channel effects are
important in the case of the scalar mesons. Using the values of
$G_S$ and $G_D$ obtained in our study of the eta mesons [1], we
found the lowest $a_0$ state at 1.13 GeV. To place the $a_0$ at
980 MeV, we increased the value of $G_{a_0}(0)$. That increase led
to the possibility of an $a_0$ condensation, which was removed by
reducing the coupling constant with increasing density. [See Fig.
8.] However, it might be preferable to accept the value of 1.13
GeV for the mass of the $a_0$ and, therefore, avoid the problem of
$a_0$ condensation. Our difficulty in this case arises since we do
not know the density dependence of the processes that move the
$a_0$ mass from our predicted value to the experimental value of
980 MeV.

In our model we see some relation between the partial restoration
of chiral symmetry and deconfinement. With reference to Fig. 2, we
see that the up (or down) quark mass drops in a roughly linear
manner with increasing density up to about 2 or 2.5 times nuclear
matter density. With the reduction of the magnitude of the
confining field, as seen in Fig. 5, the combined effect of the
smaller confining field and reduced quark mass values leads to
deconfinement at about 1.8 $\rho_{NM}$. For the $f_0$, $K$ and
$K_0^*$ mesons, the reduction of the mass of the quarks is less
important, since these mesons have either one strange quark ($K$,
$K_0^*$) or an $\bar ss$ component ($f_0$). However, deconfinement
still takes place at about $\rho=1.8\rho_{NM}$ for the mesons.

In future work we will study the dependence of the deconfinement
process on both temperature and density. In addition, it would be
desirable to have some understanding of the mechanism by which the
increased matter density modifies the confining interaction.


\vspace{1.5cm}
\noindent$\textbf{References}$\\[-2cm]


\end{document}